\documentclass[a4paper,11pt]{article}
\usepackage{pos}
\usepackage[]{graphicx}

\newcommand{\pr}{\vec{P}}
\newcommand{\pB}{\vec{p}_B}
\newcommand{\Q}{\vec{q}}

\newcommand{\tr}{\Delta t}
\newcommand{\tj}{t_J}
\newcommand{\tp}{0}

\title{A lattice QCD study of the $B \to \pi\pi \ell \bar{\nu}$ transition}

\author*[a,b]{Luka Leskovec}
\author[c]{Stefan Meinel}
\author[d]{Marcus Petschlies}
\author[e]{John Negele}
\author[f]{Srijit Paul}
\author[e]{Andrew Pochinsky}
\author[g]{Gumaro Rendon}

\affiliation[a]{Jozef Stefan Institute, Ljubljana, Slovenia}
\affiliation[b]{Faculty of Mathematics and Physics, University of Ljubljana, Ljubljana, Slovenia}
\affiliation[c]{Department of Physics, University of Arizona, Tucson, AZ 85721, USA}
\affiliation[d]{Helmholtz-Institut f\"ur Strahlen- und Kernphysik, Rheinische Friedrich-Wilhelms-Universit\"at Bonn, Nußallee 14-16, 53115 Bonn, Germany}
\affiliation[e]{Center for Theoretical Physics, Massachusetts Institute of Technology, Cambridge, MA 02139, USA}
\affiliation[f]{School of Physics and Astronomy, University of Edinburgh, Edinburgh EH9 3FD, UK}
\affiliation[g]{Zapata Computing,100 Federal St FL 20, Boston, MA 02110, United States}

\emailAdd{luka.leskovec@ijs.si}

\abstract{
  $V_{ub}$ is the smallest and least known of all CKM matrix elements; the community currently determines its magnitude primarily through the exclusive process $B\to\pi\ell\bar{\nu}$. Here we present our progress toward a lattice QCD determination of the $V_{ub}$ matrix element from a novel transition -- the $B\to\pi\pi\ell\bar{\nu}$ process, where the $\pi\pi$ system is in a $P$ wave and scattering features the $\rho(770)$ resonance as an enhancement. We perform our calculation on $N_f=2+1$ isotropic clover fermions on a lattice of $L\approx 3.6$ fm and a pion mass of $\approx 320$ MeV; for the $b$-quark we use the anisotropic clover action. After a brief overview of the theoretical framework, we will discuss some preliminary results.
}

\FullConference{%
The 39th International Symposium on Lattice Field Theory,\\
8th-13th August, 2022,\\
Rheinische Friedrich-Wilhelms-Universität Bonn, Bonn, Germany
}

\begin{document}
\maketitle

\section{Introduction}


$V_{ub}$ is the smallest and least known of all Cabibbo-Kobayashi-Maskawa (CKM) matrix elements. The community currently determines its magnitude primarily through the exclusive process $B\to\pi\ell\bar{\nu}$, although the purely leptonic $B\to \tau\nu$ and fully inclusive $B\to X_u \ell\bar{\nu}$ semileptonic decay also contribute \cite{ParticleDataGroup:2022pth}. There is, however, a puzzle regarding the determination of $V_{ub}$ - the determinations of $|V_{ub}|$ from $B\to X_u \ell\bar{\nu}$ are in tension\footnote{Recent more conservative error estimates for the inclusive determination have reduced the tension \cite{ParticleDataGroup:2022pth}.} with those using the exclusive $B\to\pi\ell\bar{\nu}$ decay rate. To better understand the tension, an additional exclusive channel to determine $V_{ub}$ would be beneficial. Such a channel is $B\to \rho(\to \pi\pi)\ell\bar{\nu}$ where the $\rho(770)$ resonance is present, which in addition to opening a new determination of $V_{ub}$, also provides complementary constraints on right-handed $b\to u$ currents for beyond the standard model physics \cite{Bernlochner:2014ova}. Experimental data for this channel are available from Babar, Belle, and Belle II \cite{delAmoSanchez:2010af, Sibidanov:2013rkk, Belle-II:2022fsw}; however, the relevant hadronic matrix elements from theory are not yet known to sufficient precision.\\

Previous lattice calculations of the $B\to \rho(\to \pi\pi)\ell \bar{\nu}$ process were done in the quenched approximation and assumed the $\rho$ resonance to be stable under the strong interaction \cite{Bowler:2004zb,Flynn:2008zr}. Here we present our preliminary results for the $B\to \rho(\to \pi\pi)\ell\bar{\nu}$ transition from lattice QCD; we perform our calculation at a pion mass where the $\rho$ appears as a resonance. To take care of the finite-volume normalization, we use Lellouch-L\"uscher factors in our analysis, enabling us to determine the transition amplitude in a range of $\pi\pi$ invariant masses, $E^\star$, and momentum transfers $q^2$. This work is the extension of our previous $\pi\gamma\to\pi\pi$ study \cite{Alexandrou:2018jbt}.

\section{Gauge Ensemble}

\begin{figure}[htb!]
 \centering
 \includegraphics[width=0.46\textwidth]{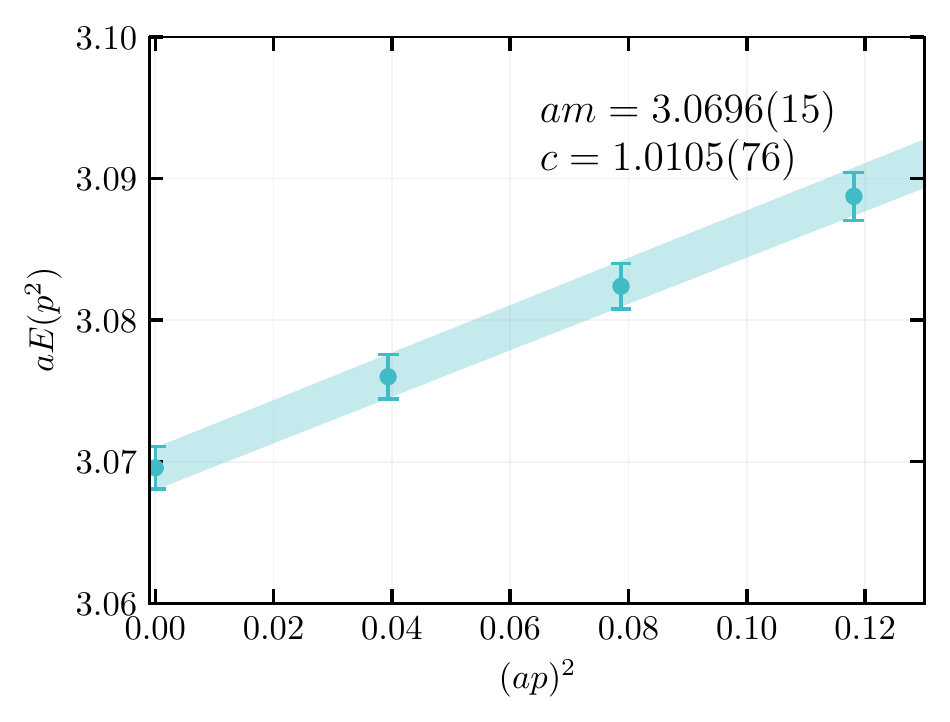}%
 \caption{The $B$-meson dispersion relation. \label{fig:Bmesondisp}}
\end{figure}

We present preliminary results on a single gauge field ensemble with $N_f=2+1$ clover Wilson fermions whose quark masses correspond to $m_\pi\approx 320$ MeV. The lattice spacing is approximately $a=0.114$ fm, and the lattice volume is $N_L^3 \times N_t = 32^3 \times 96$. The pion dispersion relation is shown in Fig.~1 of Ref.~\cite{Alexandrou:2017mpi}. For the $b$-quark, we use an anisotropic action \cite{El-Khadra:1996wdx,Chen:2000ej}, in which we tune the quark mass and anisotropy parameters to match the $B_s$ meson rest and kinetic mass. This gives the $B$-meson dispersion shown in Fig.~\ref{fig:Bmesondisp}. 

\section{The $3$-point correlation functions}

The $3$-point correlation functions definition is
\begin{align}
\label{eq:3pt}
C_{3}^i = \langle \Omega |O_i(\tr;\pr,\Lambda, r) J(\tj;\Q,\mu) B^{\dagger}(\tp;\pB)| \Omega \rangle,
\end{align}
where the source is at timeslice $0$, the current $J(\tj;\Q,\mu)$ with momentum $\Q$ at timeslice $\tj$ and the sink at timeslice $\tr$. The source part of the $3$-point correlation functions consists of a single $B$-meson interpolator, $B(\tp;\pB)=\sum_{\vec{x}} e^{i\pB \cdot \vec{x}}\bar{q}\gamma_5 b (\vec{x})$ with momentum $\pB$. For the sink part, we use three or four interpolators $O_i(\tr;\pr,\Lambda,r)$ built from either one- or two-hadron operators in the irreducible representation $\Lambda$ and row $r$. Here $\pr$ is the total momentum of the two-hadron system, projected to the irreducible representation $\Lambda$ of the Little Group defined by $\pr$. The current insertion, $J(\tj;\Q,\mu)$, is $O(a)$ improved through:
\begin{align}
\label{eq:improved}
J(\tj;\Q,\mu) = \sqrt{Z^{u}Z^{b}} (\bar{u}\Gamma b + d^{(b)} \bar{u} \Gamma \gamma^i \nabla_i b),
\end{align}
where $\Gamma$ is either $\gamma^\mu$ or $\gamma^\mu\gamma_5$, $d^{(b)}$ is the improvement coefficient, and $Z^{f}$ are renormalization coefficients of the flavor-conserving temporal vector current for quark flavor $f$. 
\begin{figure}[htb!]
 \centering
 \includegraphics[width=0.42\textwidth]{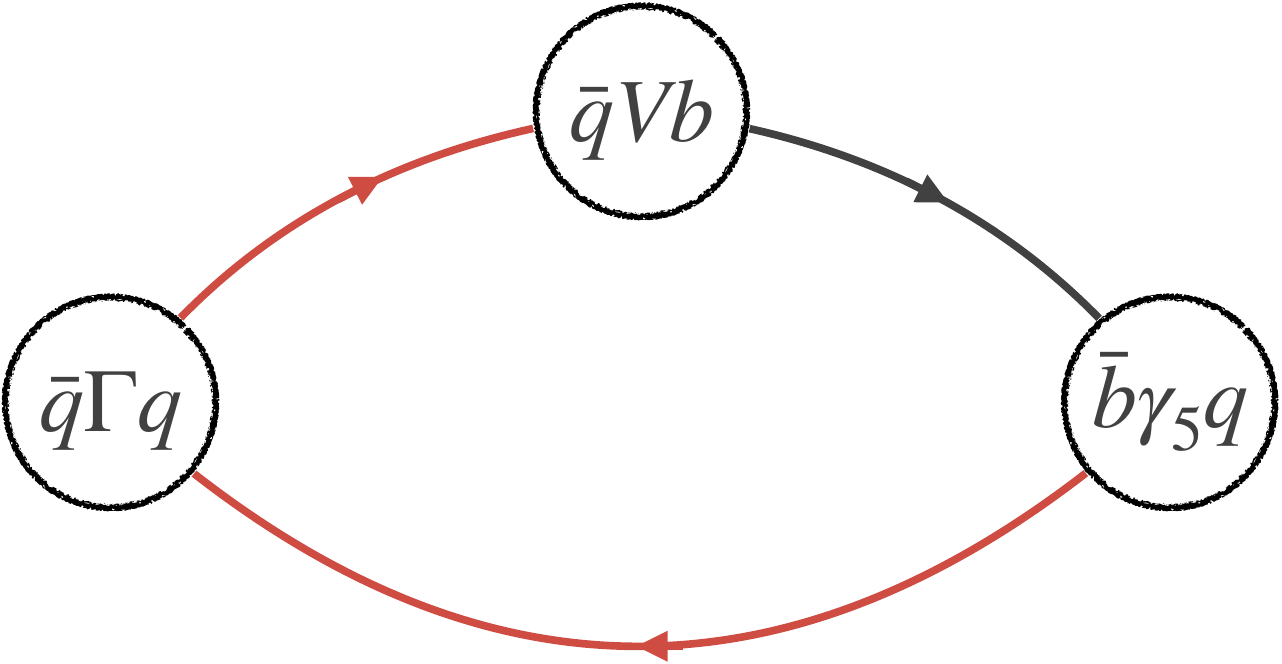}%
 \hspace{1.0cm}
 \includegraphics[width=0.42\textwidth]{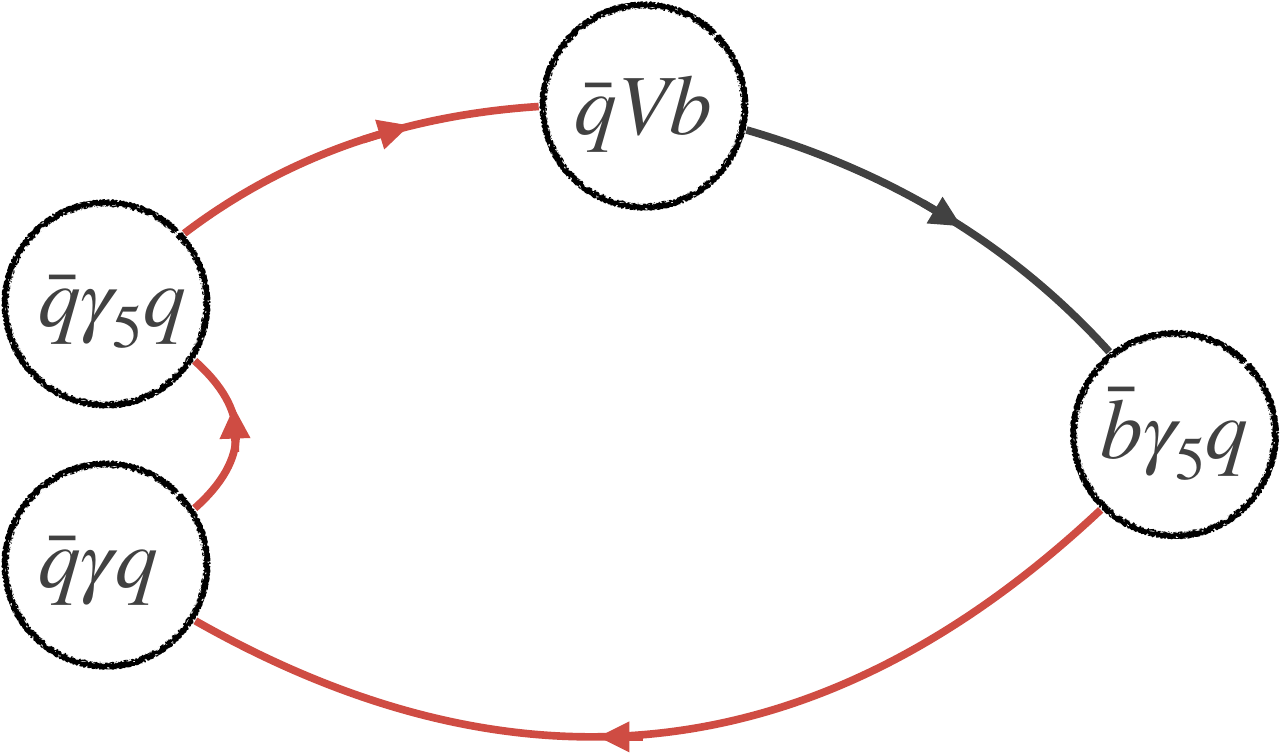}
 \caption{The Wick contractions relevant to the $B\to\pi\pi\ell\bar{\nu}$ transition. Left is the Wick contraction for the single-hadron sink operator, while the right is the Wick contraction for the two-hadron sink operator. We project both Wick contractions to $I=1$, $I_z=0$ in the sink. \label{fig:wick}}
\end{figure}
To determine the $3$-point correlation functions, we evaluate the Wick contractions shown in Fig.~\ref{fig:wick}; Fig.~\ref{fig:3pt} shows an example of the $3$-point correlation functions in the $B_3$ representation of $\vec{P}=\frac{2\pi}{L}[0,1,1]$.
\begin{figure}[htb!]
 \centering
 \includegraphics[width=0.75\textwidth]{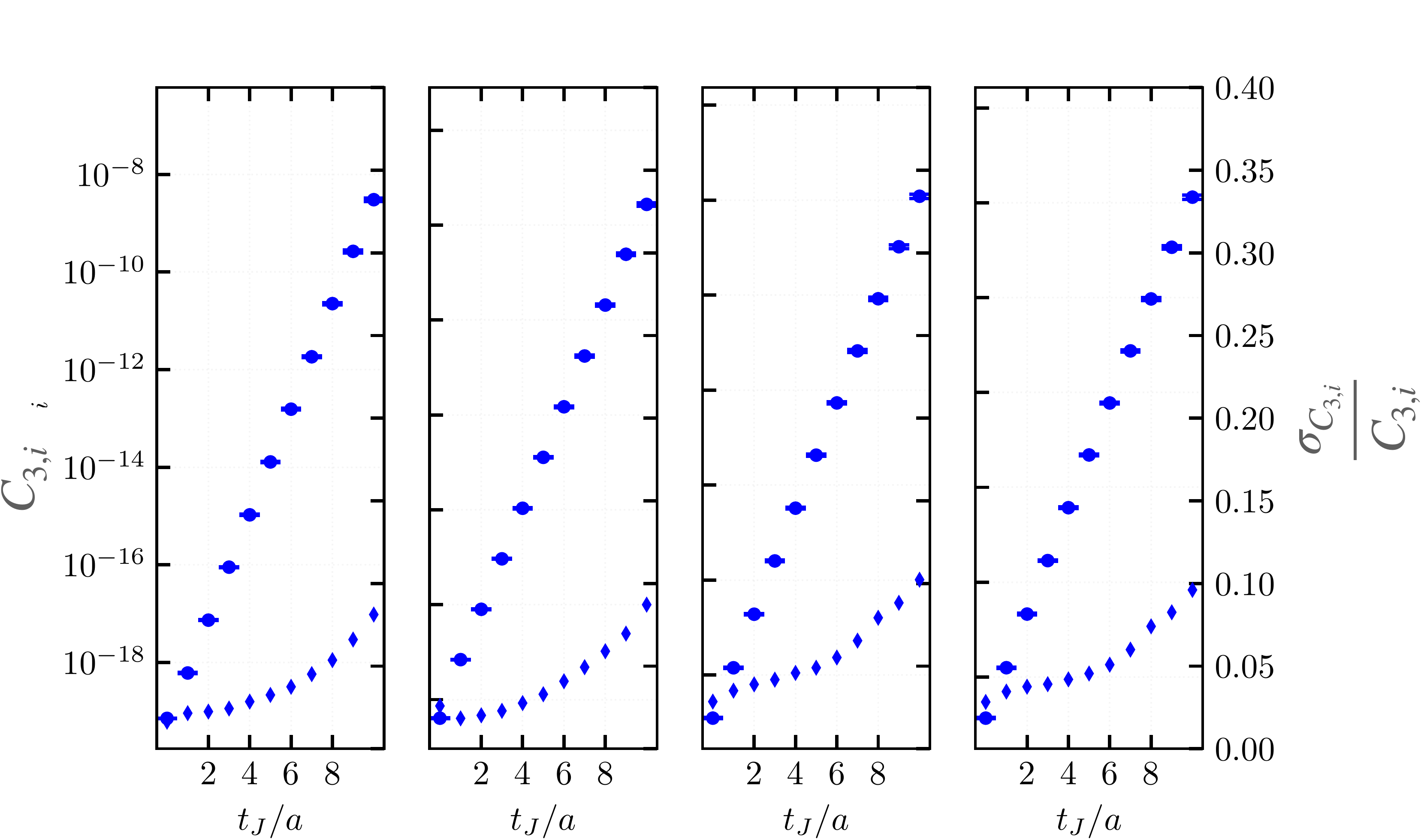}%
 \caption{Examples of $3$-point correlation functions are shown with filled circles; the left-most panel shows the $O_{i=1}=\bar{q}\gamma_i q$ operator, the second-to-left-most shows the $O_{i=2}=\bar{q}\gamma_t \gamma_i q$ operator, the second-to-right-most shows $O_{i=3}=\pi(\vec{p}_1)\pi(\vec{p}_2)$ with $|\vec{p}_1|=0$, $|\vec{p}_2|=\frac{2\pi}{L}\sqrt{2}$, and the right-most shows $O_{i=4}=\pi(\vec{p}_1)\pi(\vec{p}_2)$ with $|\vec{p}_1|=\frac{2\pi}{L}$, $|\vec{p}_2|=\frac{2\pi}{L}\sqrt{3}$. The relative uncertainties, $\frac{\sigma_{C_{3,i}}}{\bar{C}_{3,i}}$, are shown as filled diamonds. \label{fig:3pt}}
\end{figure}
The first two panels employ one-hadron, $\bar{q}\gamma_i q$ and $\bar{q}\gamma_t\gamma_i q$ sink operators, while the last two panels show $3$-point correlation functions with two-hadron operators at the sink. To construct $3$-point functions with dominant overlap to a single finite-volume state, we construct a linear combination $C_{3}^n$ of the four $3$-point correlation functions $C_{3}^i$ with coefficients $u_i^n$ taken as the $n$-th state generalized eigenvector of the GEVP analysis \cite{Alexandrou:2017mpi}, $C_{3}^n = u_i^n C_{3}^i$.
The projection leads to an optimized correlation function that dominantly overlaps with a single finite-volume state:
\begin{align}
\label{eq:3ptproj} 
C_{3}^n & = \langle n, \Lambda, \vec{P} | J | B, \pB \rangle \langle B, \pB | O_B | \Omega \rangle \frac{ e^{-E_n (\tr - \tj)} e^{-E_0^B (\tj - \tp)} }{2E_n 2E_0^B} + \text{excited state cont.},
\end{align}
where $\langle n, \Lambda, \vec{P} | J | B, \pB \rangle$ is the sought-for matrix element, and the "excited state cont." are similar products with matrix elements involving excited states of the source and sink irreducible representations. These differ from the desired matrix element in their size and temporal dependence. If we multiply the leading-order time dependence out of $C_{3}^n$, we can fit the matrix elements with models where we can consider the source, sink, or both sides of excited state contaminations. We show an example of such a matrix element in Fig.~\ref{fig:3ptproj}, where the dots with uncertainties represent the lattice data, the light-shaded region the fit of the full model, including excited state contamination, and the dark-shaded region the matrix element value. To determine the matrix elements, we vary the fit models and fit windows, which yields a total of $64$ matrix elements spread across different $q^2$ and $\sqrt{s}$. 

\begin{figure}[htb!]
 \centering
 \includegraphics[width=0.8\textwidth]{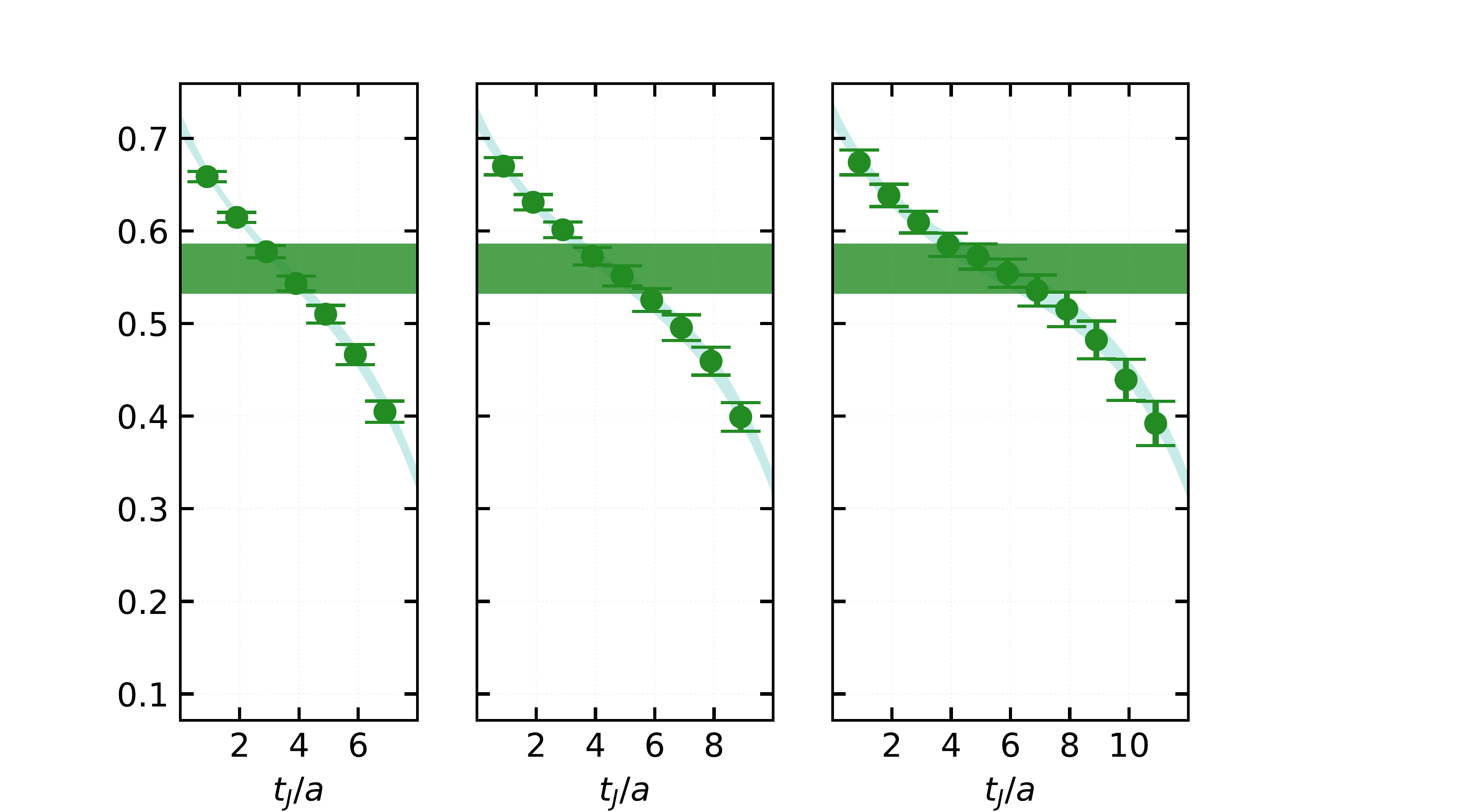}%
 \caption{An example of the state-projected $3$-point correlation function without the Lorentz symmetry factor. We factor out the leading-order temporal dependence to demonstrate the matrix element and excited state contributions. Shown is the ground state of the irreducible representation $B_3$ of $\vec{P}=\frac{2\pi}{L}[0,1,1]$ with $\pB=\frac{2\pi}{L}[0,1,1]$. The discrete data points are the lattice-determined matrix element, the light-shaded region is the full model, which includes source and sink excited-state contamination, and the dark-shaded region is the determined matrix element. \label{fig:3ptproj}}
\end{figure}

\section{Fitting the Matrix Elements}

The significant interactions in the finite-volume state (i.e., those that lead to the $\rho$ resonance) affect the normalization of the matrix elements, an effect taken into account by the Lellouch-L\"uscher factor \cite{Lellouch:2000pv,Lin:2001ek}. In these proceedings, we follow the approach of Brice\~no, Hansen, and Walker-Loud \cite{Briceno:2014uqa} and use the particular implementation discussed in Ref.~\cite{Briceno:2021xlc} to map the infinite-volume amplitudes onto the finite-volume matrix elements. For full generality, we thus do not map each finite-volume matrix element to its infinite-volume counterpart (even though this is possible in the $\pi\pi$ channel) but rather fit the finite-volume matrix elements directly. The general form of the transition amplitude ${\cal H}^\mu_{1,m_\ell}$ can be written as
\begin{align}
 \label{eq:factorization}
 {\cal H}^\mu_{1,m_\ell}(q^2,E^{\star 2}) = {\cal A}^\mu_{1,m_\ell}(q^2,E^{\star 2}) \frac{T(E^{\star 2})}{k},
\end{align}
where, for the case of the vector current, $ {\cal A}^\mu_{1,m_\ell}$ will have the following Lorentz decomposition
\begin{align}
 \label{eq:LorentzDecomp}
 {\cal A}^\mu_{1,m_\ell} = \frac{i V }{m_B+2m_\pi} \varepsilon^{\mu \nu \alpha \beta} \epsilon^{\nu\ast}(P,m_\ell) P_{\alpha} (p_{B})_{\beta}.
\end{align}
Here, V is the transition form factor, $P$ is the four-momentum of the two-hadron state, $p_{B}$ is the four-momentum of the initial $B$-meson, and $\epsilon$ is the polarization vector of the two-hadron state with $J = 1$ and third component $m_\ell$. The invariant $P\cdot P = E^{\star 2}$ denotes the $\pi\pi$ invariant mass, and the invariant $q^2$, where $q=P-p_B$, is the momentum transfer.\\

To determine the infinite-volume transition amplitude, we first pick a parameterization for $V$, set its parameters to initial values, and then obtain the finite-volume matrix elements through
\begin{align}
 \label{eq:Req}
 \langle n, \Lambda, \vec{P} | J | B, \pB \rangle = \frac{1}{\sqrt{2 E_0^B} \sqrt{2 E_n}} \;\;\sqrt{ \frac{2 E_n^{\star} }{{-\mu_0^{\star}}'} } \; V,
\end{align}
where $\mu_0^\star$ is the non-zero eigenvalue of the residue matrix $R$,
\begin{align}
 \label{eq:Rmat}
 R = 2E_n \lim_{E\to E_n}\frac{E-E_n}{F^{-1} + T},
\end{align}
and $E_n$ is the finite-volume energy corresponding to the state $\langle n, \Lambda, \vec{P} |$. In Eq.~\eqref{eq:Req}, $\mu_0^{\star'}$ is the derivative of $\mu_0^{\star}$ with respect to $E^\star$. \\

\begin{figure}[h]
  \begin{center}
  \includegraphics[width=0.9\textwidth]{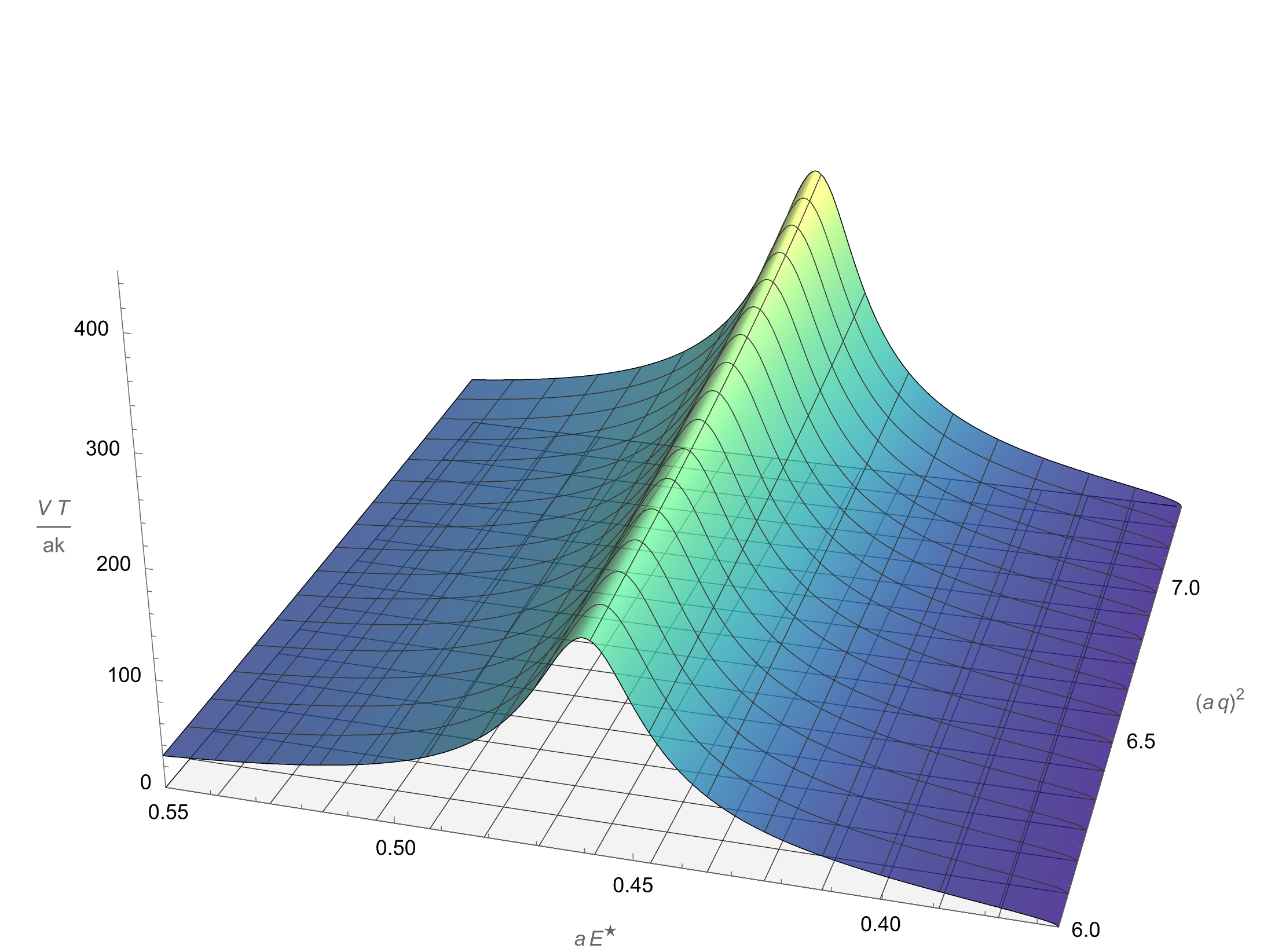}
  \end{center}
  \caption{The transition amplitude ${\cal A}(q^2,E^\star)=\frac{V T}{ak}$ of the vector current of the $B\to\pi\pi\ell\bar{\nu}$ transition in the isospin basis. Shown is the region of $q^2$ and $E^\star$, where lattice data is available.}
  \label{fig:transition}
  \end{figure}

From the model matrix element and the lattice data, we construct a $\chi^2$ function and then minimize the $\chi^2$ function for the model parameters. In these proceedings, we consider a single model of the transition amplitude, where the $\pi\pi$ scattering amplitude $T$ is the Breit-Wigner I amplitude used in Ref.~\cite{Alexandrou:2017mpi}; it fully describes the pole structure in the $E^{\star}$ variable. As such, $V$ is a smooth function of $E^{\star}$, but still has singularities in the momentum transfer, $q^2$, variable, above the semileptonic region. We parametrize $V$ using a generalization of the $z$-expansion \cite{Boyd:1994tt,Bourrely:2008za}
\begin{align}
 \label{eq:V}
 V = \frac{1}{1-\frac{q^2}{m_{B^{\star}}}} \sum_{n=0,m=0}^{n_{\rm max},m_{\rm max}} a_{n,m} z^n {\cal S}^m,
\end{align}
where ${\cal S}=\frac{E^{\star 2} - (2m_{\pi})^2}{(2m_{\pi})^2}$ and 
\begin{align}
  \label{eq:zexp}
  z = \frac{\sqrt{t_+ - q^2} - \sqrt{t_+ - t_0}}{\sqrt{t_+ - q^2} + \sqrt{t_+ - t_0}}.
 \end{align} 
Here, $t_+$  corresponds to the $B\pi$ threshold, and we use $t_0 = 6.0$ in lattice units. The $B^*$-meson pole is included explicitly as a prefactor in Eq.~\eqref{eq:V}. Our preliminary fit does not include ${\cal S}$ dependence, we used $n_{\rm max}=1$, $m_{\rm max}=0$. We show the central value of the resulting transition amplitude in Fig.~\ref{fig:transition}; as our matrix elements are in the isospin basis of the final $\pi\pi$ state, so is the transition amplitude. The fit presented used $51$ points at various $E^\star$ and $q^2$ and yields a $\chi^2/{\rm dof} = 1.4$. The corresponding parameters are $a_0=0.2405(45)$ and $a_1=-0.09(11)$.

\section{Summary}

We have presented our preliminary results for the vector form factor of the $B\to\pi\pi\ell\bar{\nu}$ transition; we plan to determine the axial-vector form factors as well. Here we presented the state-projected $3$-point correlation functions and their fits used to determine the matrix elements. Taking the Lellouch-L\"uscher factors into account, we normalize the finite-volume matrix elements that enter the global analysis of the transition amplitude. In this manner, we have determined the $B\to\pi\pi \ell\bar{\nu}$ transition amplitude in the region of large $q^2$ and $\pi\pi$ invariant mass near the $\rho(770)$ resonance. We have shown an initial fit to a subset of all our data and demonstrated the approach's viability. 

\section{Acknowledgments}

We thank Kostas Orginos, Balint Joó, Robert Edwards, and their collaborators for providing the gauge-field configurations. Computations for this work were carried out in part on (1) facilities of the USQCD Collaboration, which are funded by the Office of Science of the U.S.~Department of Energy, (2) facilities of the Leibniz Supercomputing Centre, which is funded by the Gauss Centre for Supercomputing,  (3) facilities at the National Energy Research Scientific Computing Center, a DOE Office of Science User Facility supported by the Office of Science of the U.S.~Department of Energy under Contract No.~DE-AC02-05CH1123, (4) facilities of the Extreme Science and Engineering Discovery Environment (XSEDE), which was supported by National Science Foundation grant number ACI-1548562, and (5) the Oak Ridge Leadership Computing Facility, which is a DOE Office of Science User Facility supported under Contract DE-AC05-00OR22725. L.L. acknowledges the project (J1-3034) was financially supported by the Slovenian Research Agency. S.M. is supported by the U.S. Department of Energy, Office of Science, Office of High Energy Physics under Award Number DE-SC0009913.  J.N. and A.P. acknowledge support by the U.S. Department of Energy, Office of Science, Office of Nuclear Physics under grants DE-SC-0011090 and DE-SC0018121 respectively. A.P. acknowledges support by the “Fundamental nuclear physics at the exascale and beyond” under grant DE-SC0023116.

\bibliographystyle{utphys-noitalics}
\bibliography{pos}



\end{document}